# DNN driven Speaker Independent Audio-Visual Mask Estimation for Speech Separation

*Mandar Gogate[1], Ahsan Adeel[1], Ricard Marxer[2,3], Jon Barker[3], Amir Hussain[1]*

[1]University of Stirling, UK,
[2]Université de Toulon, Aix Marseille Univ, CNRS, LIS, Marseille, France
[3]University of Sheffield, UK

`mandar.gogate@stir.ac.uk`

## Abstract

Human auditory cortex excels at selectively suppressing background noise to focus on a target speaker. The process of selective attention in the brain is known to contextually exploit the available audio and visual cues to better focus on target speaker while filtering out other noises. In this study, we propose a novel deep neural network (DNN) based audiovisual (AV) mask estimation model. The proposed AV mask estimation model contextually integrates the temporal dynamics of both audio and noise-immune visual features for improved mask estimation and speech separation. For optimal AV features extraction and ideal binary mask (IBM) estimation, a hybrid DNN architecture is exploited to leverages the complementary strengths of a stacked long short term memory (LSTM) and convolution LSTM network. The comparative simulation results in terms of speech quality and intelligibility demonstrate significant performance improvement of our proposed AV mask estimation model as compared to audio-only and visual-only mask estimation approaches for both speaker dependent and independent scenarios.

**Index Terms**: Speech Separation, Binary Mask Estimation, Deep Neural Network, Speech Enhancement

## 1. Introduction

Speech separation has received much attention in recent years due to its application in a wide range of real-world problems, ranging from automatic speech and speaker recognition, voice activity detection, to signal to noise ratio (SNR) estimation and noise reduction in hearing aids [1, 2]. In the literature, there are two widely used speech separation approaches: (1) Statistical model based background noise estimation such as spectral subtraction, Wiener filtering, and linear minimum mean square error (2) Computational auditory scene analysis (CASA) inspired by the perceptual principles of auditory scene analysis. It is well known that the statistical methods remain deficient in achieving enhanced speech intelligibility, since it introduces distortion such as musical noise. In contrast, CASA has shown to be effective in both stationary and non-stationary noises [3]. In CASA, speech is separated by applying a spectral mask to the time-frequency (T-F) representation of a noisy speech. The idea is to suppress the noise-dominant regions where background noise is stronger than target speech. The IBM is defined from the T-F representation of a speech mixture as follows:

$$IBM(t,f) = \begin{cases} 1 & \text{if } SNR(t,f) > \text{LC} \\ 0 & otherwise. \end{cases} \quad (1)$$

The IBM assigns unit value to a T-F unit if the local SNR within unit exceeds the local criterion (LC), and 0 otherwise.

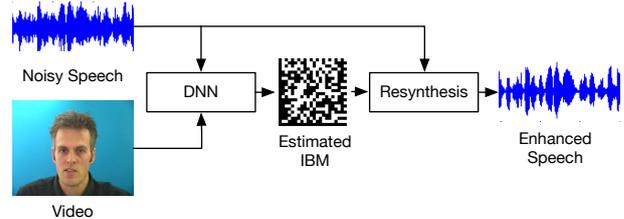

Figure 1: *System overview: Audio-Visual Mask Estimation for Speech Separation*

The IBM has shown to improve the overall speech quality and intelligibility for normal-hearing and hearing-impaired listeners [4, 5]. In real world scenarios, IBM cannot be calculated using equation 1. However, the problem can be modelled as a data-driven binary classification problem that uses noisy speech to estimate the IBM.

In the literature, extensive research has been carried out to develop speech separation methods for speech recognition [6, 7]. Researchers have proposed several different speech separation models such as parametric mask estimation methods [8, 9, 10], neural network based mask estimation methods [11, 12], and novel loss functions [13]. However, limited work has been conducted to develop robust speaker independent audio-visual speech separation models to perform enhancement. The few attempts to address this problem have been restricted to speaker dependent scenarios. In [13], audio and visual features are first concatenated into a single vector. The concatenated vector is then used to train a non-causal speaker-dependent DNN with a perceptually motivated loss function inspired by the hit minus false-alarm (HIT-FA) rate. In addition, [14] proposed an audio-based blind source separation to extract the target speaker from speech mixture. The permutation and scaling ambiguities present in the estimated signal are corrected using visual speech information.

In supervised learning, generalisation to unseen data is a critical issue. One of the major issues of supervised speech separation are noise and speaker generalisation. It has been shown that given enough training noises, a DNN generalises well to unseen ones [15]. However, the generalisation capability to unseen speakers with an unknown noise remains a challenging task. In addition, it has been shown that, the concatenation of raw unimodal features into a single vector degrades the overall performance of the supervised learning system [16].

In this study, we propose a unified model that separates speech of an unseen speaker from an unknown noise. The developed hybrid DNN model integrated a stacked LSTM and Con-volutional LSTM network for optimised AV mask estimation, taking into account the temporal dynamics of both audio and visual data. The developed model exploits the long-term contextual dependencies to better focus on the target speaker. An overview of our proposed AV driven mask estimation model is shown in Figure 1. The hybrid DNN model effectively learns the correlation between noisy AV features and IBM in order to produce an estimate of noise and speech dominant regions. The estimated IBM retains the speech dominant T-F units and suppresses the noise dominant T-F units. Finally, the enhanced speech is resynthesised by combining processed signal across frequency channels.

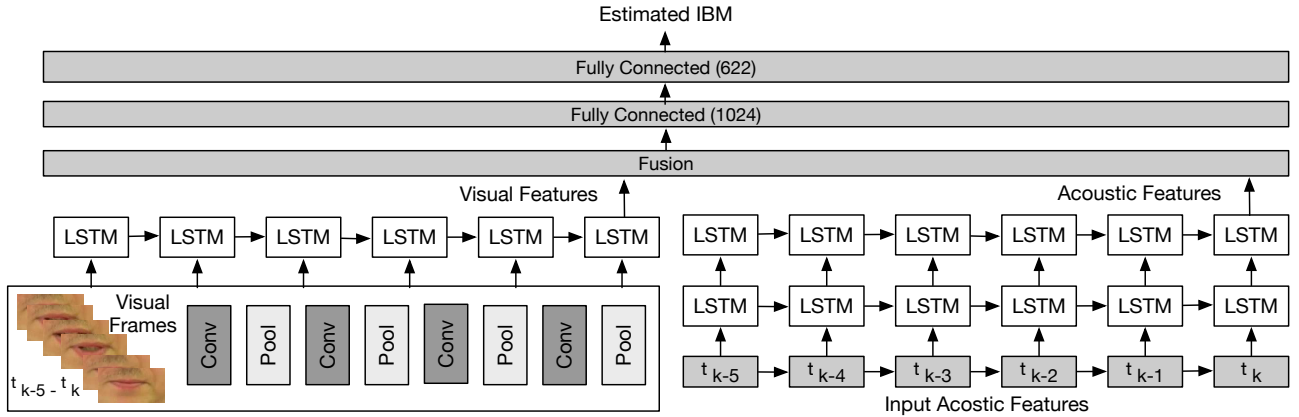

Figure 2: *Proposed Multimodal Mask Estimation Model*

The rest of the paper is organised as follows: Section 2 presents the dataset and preprocessing. Section 3 presents the proposed DNN driven AV mask estimation. Section 4 explains the experimental results. Finally, Section 5 concludes this work.

## 2. AV Dataset and pre-processing

### 2.1. Dataset

In our experiments, the widely used benchmark Grid Corpus [17] is used. For preliminary analysis, 5 speakers are considered (speaker 1, 6, 7, 15 and 26) each reciting around 1000 utterances, and each sentence consists of a six word sequence of the form indicated in Table 1. The clean Grid utterances are mixed with random highly non-stationary noises (bus, cafe, street, pedestrian) from 3rd CHiME Challenge (CHiME3)[18] for different SNR levels ranging from -12dB to 6dB with a step size of 6dB. It is to be noted that, all the models used were SNR-independent: all the utterances mixed at all SNR levels were employed for training and testing.

Specifically, two different models were trained: (1) speaker dependent (2) speaker independent

1. **Speaker Dependent**: In speaker dependent analysis, 3000 utterances from all 5 speakers were used for training and validation of the DNN model, with a validation split of 20%. The trained network was evaluated on a test set of 2 seen speakers (15 and 26).
2. **Speaker Independent**: In speaker independent analysis, 3000 utterances from 3 speakers (1, 6 and 7) were used for training and validation of the DNN model, with a validation split of 20%. The trained network was evaluated on a test set of 2 unseen speakers (15 and 26).

Table 1: *GRID sentence grammar*

| command | colour | preposition | letter | digit | adverb |
|---|---|---|---|---|---|
| bin | blue | at | A-Z | 1-9 | again |
| lay | green | by | minus W | zero | now |
| place | red | in | | | please |
| set | white | with | | | soon |

### 2.2. Pre-processing

**Audio**: The noisy audio signal is sampled at 16kHz and segmented into N 80 ms frames with 1200 samples per frame and 25% increment rate. A Short Time Fourier Transform (STFT) and hamming window is applied to produce 622-bin power spectrum.

**Video**: The speakers' lip images are extracted out of the GRID Corpus videos (recorded at 25 frames per second) using a Viola-Jones lip detector [19] and an object tracker [20]. A region of 92 x 50 was selected around the lip centre. In addition, the extracted lip sequences were upsampled by 3 to match the 75 vectors per second (VPS) rate for audio files.

## 3. DNN driven AV Mask Estimation

This section describes the network architecture, depicted in Figure 2. The network shown ingests preprocessed audio and visual features of time instance $t_k$, $t_{k-1}$, ... ,$t_{k-5}$ (k is the current time instance and 5 is the number of prior visual frames).

The network consists of three main components: (1) Audio Feature Extraction (2) Visual Feature Extraction, and (3) Multimodal Fusion.

### 3.1. Audio feature extraction

The preprocessed power spectra of time instances $t_k$ to $t_{k-5}$ were fed into a two layered stacked LSTM network consisting of 1024 cells each. A dropout of 0.2 was applied after each LSTM layer to prevent the network from overfitting the train data. The last outputs of the 2nd LSTM layer were used as acoustic features for multimodal fusion.

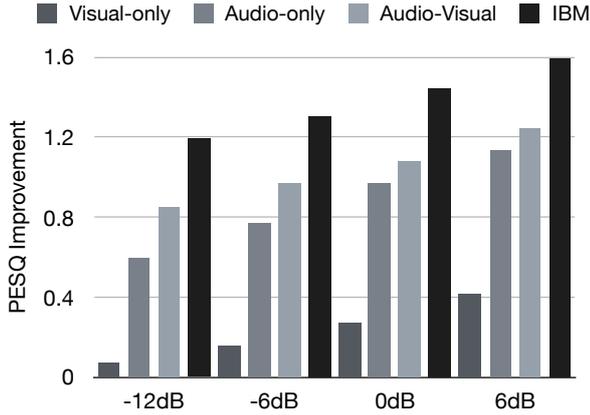

Figure 3: *PESQ improvement computed from the reconstructed speech signal using Visual-only, Audio-only, Audio-visual mask estimation methods, and IBM for unseen speakers*

### 3.2. Visual feature extraction

The extracted and preprocessed lip-regions were fed into a convolution LSTM model. The model has input dimensions of W x 50 x 92, where *W* is the number of prior visual frames. Convolution filters are applied to these concatenated utterances. The CNN has a total of 8 layers: 4 convolution and 4 max pooling layers, consisting 32, 64, 64, and 128 feature maps. Moreover, each convolution uses filters of size 3 x 3 , followed by a max pooling layer with a window size of 2 x 2. The individual feature maps extracted from the last max pooling layer were fed into a LSTM network with 1024 cells. The LSTM outputs were used as visual features for multimodal fusion.

### 3.3. Multimodal fusion

The optimal features extracted from audio (1024-d) and visual (1024-d) modalities were concatenated into a single vector. The concatenated vector (2048-d) was fed into a fully connected multi-layered perceptron (MLP) network. The MLP consists of two layers having 1024 (ReLU function) and 622 (Sigmoidal function) neurons each. The multimodal fusion method exploited the complementary strengths of both early and late fusions. It is to be noted that the audio (A) only and video (V) only mask estimation models were constructed by eliminating the visual and audio feature extraction parts from the networks respectively. It is worth mentioning that no thresholding was applied and sigmoidal outputs were considered as the estimated mask.

## 4. Results

### 4.1. Experimental Setup

For audiovisual cues integration and mask estimation, deep learning based data-driven approaches are trained and validated using Tensorflow library and 4 NVIDIA Titan X Pascal GPUs with 12GB of GDDR5X memory each. A subset of dataset is used for training/validation of the neural network (80% training dataset) and rest of the data is used to test the performance of the trained neural network in unseen scenarios (20% testing dataset). The pre-processed training set of AV corpus consists of around 3000 utterance, which are further split into 2400 and 600 utterances for training and validation respectively. The dataset consists of visuals and noisy audio frames as input to the learning model and IBM as an output. Training was performed using backpropagation with the Adam optimiser [21]. The DNN performances are evaluated using classification accuracy of T-F units and perceptual evaluation of speech quality (PESQ) score.

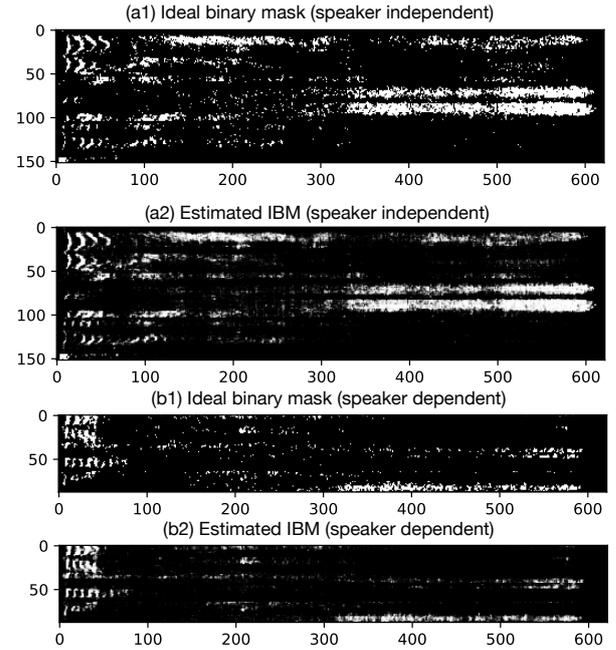

Figure 4: *Ideal binary mask and Estimated IBM by DNN, for (a) Speaker Independent (b) Speaker Dependent cases*

### 4.2. Time-Frequency Classification Results

The classification results of our proposed DNN models are summarised in Table 2. It can be seen that going from low SNRs to high SNRs, the classification accuracy decreases linearly for all deep learning models (A-only, V-only, and AV) in both speaker dependent and independent scenarios. It is mainly because of the total number of 0's and 1's for each case, where at low SNRs there are less transitions as compared to transitions at higher SNRs. Overall, AV driven mask estimation model significantly outperformed both A-only and V-only driven mask estimation models, where V-only achieved the least classification accuracy for all SNRs in both speaker dependent and independent cases. It is worth mentioning that AV driven mask estimation achieved the highest accuracy of 95.5%, 94.8%, and 92.1% at -12dB, -6dB, and 0dB SNRs respectively.

### 4.3. Objective testing

In order to objectively measure the quality of re-synthesised speech, a widely used PESQ test is used. PESQ is an mean opinion score (MOS) like evaluation, though correlation between subjective MOS test and PESQ test is not very high. However, among all objective measures, PESQ is one of the reliable methods to evaluate speech quality. The PESQ score is computed as a linear combination of the average disturbance value and the average asymmetrical disturbance values. The PESQ score ranges from -0.5 to 4.5 corresponding to low to high speech quality. The PESQ results for A-only, V-only, and AV models are presented in Table 3 for both speaker dependent and inde-pendent cases. It can be seen that in speaker independent scenario, AV significantly outperformed A-only and V-only mask estimation models, where AV model achieved the PESQ scores of 1.87, 2.05, and 2.17 at SNRs levels of -12dB, -6dB, and 0 dB respectively, as compared to 1.09, 1.23, and 1.36 achieved by V-only and 1.61, 1.84, and 2.05 achieved by A-only mask estimations. At high SNR (6dB), AV mask estimation achieved PESQ score of 2.34, as compared to 2.23 and 1.51 achieved by A-only and V-only driven mask estimations respectively. The overall PESQ improvement as compared to noisy audio is depicted in Figure 3, where AV-driven mask estimation significantly outperformed the A-only and V-only driven mask estimation models, and achieved near optimal performance (close to IBM). Figure 4 presents the IBM and estimated IBM by DNN of a randomly selected utterance, where the quality of mask estimation is apparent (i.e. close to the ideal).

Table 2: *Mask estimation results (T-F classification accuracy): Comparison of audio-only, visual-only, and audio-visual mask estimation models*

| SNR | Speaker Dependent | | | Speaker Independent | | |
|---|---|---|---|---|---|---|
| | Visual-only | Audio-only | Audio-Visual | Visual-only | Audio-only | Audio-Visual |
| -12dB | 92.8 | 96.7 | 97.5 | 92.1 | 92.1 | 95.5 |
| -6dB | 91.6 | 95.5 | 96.9 | 90.8 | 90.1 | 94.8 |
| 0dB | 90.5 | 94.2 | 96.1 | 89.7 | 88.6 | 92.1 |
| 6dB | 86.5 | 93.7 | 95.6 | 84.9 | 85.2 | 89.9 |

Table 3: *PESQ scores: Comparison of audio-only, visual-only, and audio-visual mask estimation models with IBM and noisy speech*

| SNR | Speaker Dependent | | | Speaker Independent | | | Noisy | IBM |
|---|---|---|---|---|---|---|---|---|
| | Visual-only | Audio-only | Audio-Visual | Visual-only | Audio-only | Audio-Visual | | |
| -12dB | 1.21 | 1.72 | 2.18 | 1.09 | 1.61 | 1.87 | 1.018 | 2.21 |
| -6dB | 1.30 | 1.95 | 2.34 | 1.23 | 1.84 | 2.05 | 1.07 | 2.38 |
| 0dB | 1.52 | 2.14 | 2.42 | 1.36 | 2.05 | 2.17 | 1.08 | 2.53 |
| 6dB | 1.73 | 2.33 | 2.53 | 1.51 | 2.23 | 2.34 | 1.09 | 2.69 |

## 5. Conclusions

This paper presented a novel DNN based AV mask estimation model that contextually integrates and exploits the temporal dynamics of both audio and visual features for enhanced mask estimation and speech separation. The multimodal hybrid DNN architecture exploited the complementary strengths of both early and late fusions using LSTM and CLSTM networks. The performance evaluation in terms of mask estimation accuracy, speech quality, and speech intelligibility revealed significant performance improvement of our proposed AV model as compared to both A-only and V-only driven mask estimation models. Specifically, AV driven mask estimation achieved the highest accuracy of 95.5%, 94.8%, and 92.1% at -12dB, -6dB, and 0dB SNRs respectively in speaker independent scenarios. The mask estimation improvement also reflected in PESQ speech quality evaluation, where AV-driven mask estimation significantly outperformed the A-only and V-only driven mask estimation models, and achieved near optimal speech enhancement performance. It is worth mentioning that the Grid corpus is very regular and could help achieving higher accuracy. In future, we intend to investigate the generalisation capability of our proposed DNN model with more other more challenging audiovisual corpora.

## 6. Acknowledgements

This work was supported by the UK Engineering and Physical Sciences Research Council (EPSRC) Grant No. EP/M026981/1. We gratefully acknowledge the support of Prof. Hadi Larijani from Glasgow Caledonian University and NVIDIA Corporation for providing access to the Titan X Pascal GPUs for this research.

## 7. References

[1] A. Narayanan and D. Wang, "Investigation of speech separation as a front-end for noise robust speech recognition," *IEEE/ACM Transactions on Audio, Speech, and Language Processing*, vol. 22, no. 4, pp. 826–835, 2014.

[2] H. Kayser, C. Spille, D. Marquardt, and B. T. Meyer, "Improving automatic speech recognition in spatially-aware hearing aids," in *Sixteenth Annual Conference of the International Speech Communication Association*, 2015.

[3] J. Chen and D. Wang, "Dnn based mask estimation for supervised speech separation," in *Audio source separation*. Springer, 2018, pp. 207–235.

[4] M. Ahmadi, V. L. Gross, and D. G. Sinex, "Perceptual learning for speech in noise after application of binary time-frequency masks," *The Journal of the Acoustical Society of America*, vol. 133, no. 3, pp. 1687–1692, 2013.

[5] D. Wang, U. Kjems, M. S. Pedersen, J. B. Boldt, and T. Lunner, "Speech intelligibility in background noise with ideal binary time-frequency masking," *The Journal of the Acoustical Society of America*, vol. 125, no. 4, pp. 2336–2347, 2009.

[6] H. Erdogan, J. R. Hershey, S. Watanabe, and J. Le Roux, "Deep recurrent networks for separation and recognition of single-channel speech in nonstationary background audio," in *New Era for Robust Speech Recognition*. Springer, 2017, pp. 165–186.

[7] A. Narayanan and D. Wang, "Ideal ratio mask estimation using deep neural networks for robust speech recognition," in *2013 IEEE International Conference on Acoustics, Speech and Signal Processing*, May 2013, pp. 7092–7096.

[8] N. Ito, S. Araki, and T. Nakatani, "Permutation-free convolutive blind source separation via full-band clustering based on frequency-independent source presence priors," in *2013 IEEE International Conference on Acoustics, Speech and Signal Processing*, 2013.

[9] N. Ito, S. Araki, T. Yoshioka, and T. Nakatani, "Relaxed disjointness based clustering for joint blind source separation and dereverberation," in *Acoustic Signal Enhancement (IWAENC), 2014 14th International Workshop on*. IEEE, 2014, pp. 268–272.

[10] D. H. T. Vu and R. Haeb-Umbach, "Blind speech separation employing directional statistics in an expectation maximization framework," in *Acoustics Speech and Signal Processing (ICASSP), 2010 IEEE International Conference on*. IEEE, 2010, pp. 241–244.

[11] J. Heymann, L. Drude, and R. Haeb-Umbach, "Neural network based spectral mask estimation for acoustic beamforming," in *Acoustics, Speech and Signal Processing (ICASSP), 2016 IEEE International Conference on*. IEEE, 2016, pp. 196–200.

[12] J. Chen and D. Wang, "Long short-term memory for speaker generalization in supervised speech separation," *The Journal of the Acoustical Society of America*, vol. 141, no. 6, pp. 4705–4714, 2017.

[13] D. Websdale and B. Milner, "A comparison of perceptually motivated loss functions for binary mask estimation in speech separation," in *Interspeech 2017, 18th Annual Conference of the International Speech Communication Association, Stockholm, Sweden, August 20-24, 2017*, 2017, pp. 2003–2007.

[14] Q. Liu, W. Wang, and P. Jackson, "Audio-visual convolutive blind source separation," in *Sensor Signal Processing for Defence (SSPD 2010)*, Sept 2010, pp. 1–5.

[15] J. Chen, Y. Wang, S. E. Yoho, D. Wang, and E. W. Healy, "Large-scale training to increase speech intelligibility for hearing-impaired listeners in novel noises," *The Journal of the Acoustical Society of America*, vol. 139, no. 5, pp. 2604–2612, 2016.

[16] M. Gogate, A. Adeel, and A. Hussain, "A novel brain-inspired compression-based optimised multimodal fusion for emotion recognition," in *2017 IEEE Symposium Series on Computational Intelligence (SSCI)*, Nov 2017, pp. 1–7.

[17] M. Cooke, J. Barker, S. Cunningham, and X. Shao, "An audio-visual corpus for speech perception and automatic speech recognition," *The Journal of the Acoustical Society of America*, vol. 120, no. 5, pp. 2421–2424, 2006.

[18] J. Barker, R. Marxer, E. Vincent, and S. Watanabe, "The third chimespeech separation and recognition challenge: Dataset, task and baselines," in *Automatic Speech Recognition and Understanding (ASRU), 2015 IEEE Workshop on*. IEEE, 2015, pp. 504–511.

[19] P. Viola and M. Jones, "Rapid object detection using a boosted cascade of simple features," in *Computer Vision and Pattern Recognition, 2001. CVPR 2001. Proceedings of the 2001 IEEE Computer Society Conference on*, vol. 1. IEEE, 2001, pp. I–511.

[20] D. A. Ross, J. Lim, R.-S. Lin, and M.-H. Yang, "Incremental learning for robust visual tracking," *International Journal of Computer Vision*, vol. 77, no. 1-3, pp. 125–141, 2008.

[21] D. P. Kingma and J. Ba, "Adam: A method for stochastic optimization," *arXiv preprint arXiv:1412.6980*, 2014.